\documentclass[showpacs,twocolumn]{revtex4}
\usepackage[dvips]{epsfig}

\newcommand{\be}{\begin{equation}}
\newcommand{\ee}{\end{equation}}
\newcommand{\cT}{{\cal T}}
\newcommand{\rgl}{\rangle}
\newcommand{\lgl}{\langle}
\newcommand{\prt}{\partial}
\begin{document}

\title{On tumor development: fractional transport approach }

\author{A. Iomin$^*$ and S. Dorfman}
\affiliation{Department of Physics, Technion - Israel Institute of Technology, Haifa, 32000, Israel }
\author{L. Dorfman }
\affiliation{Medical School and Department of Microbiology and Immunology,
Faculty of Health Science, 
Ben--Gurion University of Negev,  IL-84105
Beer Sheva, Israel}

\date{\today}
\begin{abstract}
A growth of malignant neoplasm is
considered as a fractional transport approach. We suggested that the main process of the tumor
development through a lymphatic net is fractional transport of
cells. In the framework of this fractional kinetics we were able
to show that the mean size of main growth is due to subdiffusion,
while the appearance of metaphases is determined by
superdiffusion.
\end{abstract}

\maketitle

\section{Introduction}

Application of fractional calculus in biology is mainly used to
describe an activity of living system(see recent reviews
\cite{from_hillfer}). Our purpose is to describe a 
tumor growth process by means of cell transport
inside a fractional network. In the present consideration  we propose  a link between
tumor spread  and fractional transport, whose
mathematical apparatus is well established
\cite{hillfer,oldham,klafter} 
We propose simplified
mathematical models using heuristic arguments on tumor growth. On
the basis of statement, we are citing  \cite{textbook}: "Lymphatic spread is more typical of
carcinomas whereas the hematogenous route is favored by sarcomas.
There are numerous interconnections, however, between the
lymphatic and vascular systems, and so all forms of cancer may
disseminate through either or both systems." - we
focus on a possible mechanism of this growth as a result of
fractional transport of cancer cells along lymphatic net system,
which has a fractal Hausdorff dimension \cite{from_hillfer}. Using
a simplified approach of fractional transport, we are also 
suggesting a possible answer
on a question how neoplasm cells appear arbitrary far from a
main tumor. The tumor growth is a complicated process which
consists of consequences of various and different phases. This phases
can be considered as  independent processes with different time
duration. The last depends on various  environmental conditions as well.

The growth of malignant neoplasm can be described as a two-step
process \cite{textbook}. The first phase is the growth and
differentiation of the primary tumor and the second phase can be
described as a metastatic phase, during which there are many
changes in the primary tumor's cells that lead to the metastasis. The
difference between the methastatic cells and the cells of the
primary tumor includes, for example, a different expression of
adhesion molecules such as E-cadherins, catenins etc.
\cite{difmolecules}. Those differences are discussed in many
studies, see e.g. \cite {differences}, according to which after
certain proliferation of the primary tumor the metastatic stage
begins - the cells start to express different molecules. Those
cells separate from the primary tumor - a process is named
metastasis.

The metastasis cells can be spread throughout the lymphatic
net system or throughout the blood stream 
(vascular system) \cite{textbook}. Apparently,
there are cytokines which support the lymphatic proliferation and
encourage the directional movement toward lymphatic system
\cite{lympha}. There are tumors which express different specific
chemokine receptors and they are more likely to spread through the
lymphatic system.

In the present  paper we suggest a simplified scheme of tumor spread
along the lymphatic net. In our scheme
we are introducing two - step process, where
the first one is a biological process of cells fission. The second
phase is cells transport itself inside lymphatic net structure. We primary
focus on the second phase, supposing that the
main size of the tumor development is due to fractional transport, and present
the two times tumor development model (TTT-model). 

\section{Main assumptions for the two times tumor development model}

In order to construct the TTT-model for the tumor development description
we are introducing a number of assumptions which are essential for our 
approach:

\noindent{\em (a)} We suppose, that there are two time scales $ \cT_f
$ and $\cT_t$. The first one corresponds to the volume grow due to
proliferation of cells by fission. During this time  cells are
strongly interacted \cite{tempia}. There is no cell transport, 
and only the fed
cells can proliferate. The duration of $\cT_f$ could be arbitrary
large and it can reach $10^7-10^8 sec$ \cite{textbook}. The second time $\cT_t$ 
corresponds
to cell transport. We suppose  that during this time there are
enough  cells without fission, and the interaction between tumor
cells is rather small. Hence the cells form an initial packet of free
particles, which spread along the lymphatic net. 

\noindent {\em (b)} The lymphatic net is a system of organic tubes with
complicated crosses and arbitrary (random) number of entrances at
nodes. It fulfilled by lymphatic liquid with laminar flow.  Citing \cite{textbook,tmph}
``About 100 milliliters per our of lymph flows through the thoratic duct 
of a resting human and approximately another 20 milliliters flows into the 
circulation each hour through other channels, making a total estimated 
lymph flow of about 120{\em ml/hr}, that is between 2 and 3 liters per 
day''.  The Hausdorff dimension 
of this geometry is fractal $ d_f<3 $ (3 is embedding dimension). Since the 
geometrical  dimension of the lymphatic net is fractional, the neoplasm 
development corresponds to anomalous diffusion.

\noindent {\em (c)} The tumor spreading  process consists of the
following time consequences
\be\label{c1}
 \cT_f(1)\cT_t(2)\cT_f(3)\dots \, .
\ee
There are different realizations of 
this chain of times, due to different duration of $\cT_f(i)$ and $\cT_t(i)$,
where $i=1, 2, \dots$ and it means simply a step--number of the 
process.
It should be underlined that the processes with different steps $i$ are 
absolutely independent and there is possible any realization of the time 
duration $\cT_t(i)$. Therefore, one can introduce the probability 
distribution function (PDF) for $\cT_t(i)$. 
 As a result of
these realizations, there are few scenarios of the tumor development.
This process depends on  the rate of the tumor cell spreading.
If transport is subdiffusive, that is typical for the lymphatic spread of 
carcinomas \cite{textbook},  it
can be described in the framework of continuous time random walk
(CTRW) \cite{shlesinger,klafter}. 
 The transport can be superdiffusive 
due to so-called Levy-type flights when cells may traverse
all of the lymph nodes ultimately to reach the vascular compartment via the
thoratic duct.
In this case for any anatomic localizations metastasis are possible.
It should be stressed  that an every cell (a particle)
carries with itself its own trap, and this is the principal deference from 
the standard CTRW, where traps are external with respect to the 
transporting particles.  


\section{Subdiffusive growth}

There are many possibilities to describe the cell transport as a diffusive 
process
taking place during the time scales $\cT_t(i)$ in the framework
of the fractional calculus. One of the simplest approximation for 
this process is CTRW. We used here a simplified model which neglects
proliferation of cells. It means that a number of cells participated in 
the transport is conserved.
Suppose  that $\cT_f(i)\gg\cT_t(i)$, and following the CTRW
construction (see, for example, \cite{klafter}), we believe that
during $\cT_t(i)$ a cell ``jumps'' with the jump length variance
\be\label{c2} 
\sigma^2=\int_{-\infty}^{\infty}l^2\psi(l,t)dldt 
\ee
and $\cT_f(i)$ are waiting times elapsing between two successive
jumps, where 
\be\label{c3} \cT=\int_0^{\infty}t\psi(l,t)dldt 
\ee 
is the mean waiting time. These values are characterized by a
probability distribution function (PDF) $\psi(l,t)$. If the
waiting time distribution has the following form \be\label{c4}
\tilde{\psi}(t)=\int_0^{\infty}\psi(l,t)dl\sim
1/(1+t/\cT)^{1+\alpha}, \ee with $ 0<\alpha<1 $ the behavior for a
CTRW is subdiffusive with the mean squared displacement being
\be\label{c5}
\lgl l^2(t)\rgl\sim t^{\alpha}. \ee
 This
subdiffusive relaxation process is responsible for the size of 
the primary tumor. It may be relevant if one takes
into account that during times $\cT_t(i)$ there is diffusion on
fractal lymphatic net of dimension $d_f$. The mean squared displacement in
this case is given by (\ref{c5}), where the transport exponent
$\alpha$ depends now on $d_f$. In these cases the relaxation
process is modeled by fractional diffusion equation, where 
the PDF $ \psi(l,t) $ has non-Gaussian form. The square root from 
mean squared
displacement $\lgl l^2\rgl $ is an average size of the main tumor
development.

When $\alpha>1$, the transport on the fractal lymphatic net is
superdiffusive. It may describe  a metastasis process. In this
case $\sqrt{\lgl l^2(t)\rgl} $ is an average distance between a
new tumor location and  the primary tumor after time
$t$. This time $t\sim \cT$ is a time when a cell moves long enough
without fission.

\section{Comb model}

In what follows
we consider a simple model of fractional transport of neoplasm
cells. In the framework of this model we are able to determine the
size of the tumor development.
Anomalous diffusion on a comb model is an oversimplified presentation of
diffusion on the fractal lymphatic net. The model was proposed in \cite{em1}
for drift particles in percolation systems and used for some
applications in solid state  \cite{baskin} and biophysics \cite{em3}.
The comb--like geometry shown in Fig. 1 is a toy model of a porous
diffusion medium.
We borrow
this consideration from \cite{baskin}  primary to describe subdiffusion 
of cancer cells on the lymphatic net and demonstrate a fractal nature of 
cancer cells transport as well as to determine a size of cancer growth.
 A special feature of diffusion in this model is a
possibility of motion in the $x$-direction only along the structure axis,
when $y=0$. In this case motion in the $y$--direction is a trap, where a
particle stays in the comb teeth for some time before it moves along the
structure $x$--axis.  This model approximates cell transport along the 
lymphatic net embedded in 3D sphere $(R, \theta, \phi)$. Diffusion along the 
structure axis (y=0) specifies motion along the lymphatic net leading to 
the increasing of the radial size $ R=R(t)$ of the 
tumor. Motion along the y-direction corresponds to cell transport in the  
polar and azimuthal directions. 
This motion does not lead to the increase of the radial size of the tumor
and may be considered as a trap.
 The PDF of waiting or delay times for the infinite
teeth is $\psi(t)\sim t^{-\beta}$, where $ 1<\beta<2$, such that the
mean waiting time is diverged. It follows from the presented geometry,
that the diffusion coefficient $D_{xx}$ differs from zero only when
$y=0$, namely $ D_{xx}=D_1\delta(y) $. The diffusion coefficient along the
$y$ direction is $D_{yy}=D_2$ and $ D_{xy}=D_{yx}=0$.
\begin{figure}
\begin{center}
\epsfxsize=7.6cm
\leavevmode
    \epsffile{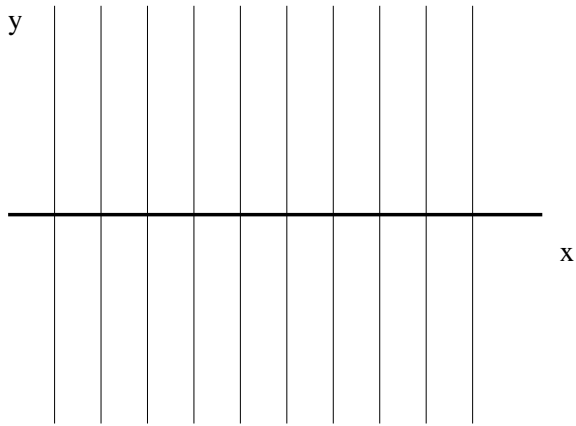}
\caption{The comb structure}
\end{center}
\end{figure}
Therefore, The Fokker--Planck (FP) equation for the Green function
$G\equiv G(x,y,t)$ reads \cite{baskin} 
\be\label{c7} 
\frac{\prt G}{\prt t} -D_1\delta(y)\frac{\prt^2 G}{\prt x^2}- D_2\frac{\prt^2
G}{\prt y^2}=\delta(x)\delta(y)\delta(t) \, . 
\ee 
In the
Laplace--Fourier $(p,k)$ space after carrying out corresponding
transformations for the $ x,t $ variables we obtain from Eq.
(\ref{c7}) the following solution \cite{baskin}
\be\label{c9}
G(k,y,p)=\frac{\exp\left[-(p/D_2)^{1/2}|y|\right]}
{2(D_2p)^{1/2}+D_1k^2} \, . 
\ee
For diffusion along the structure
axis one can infer the Green function $c(x,t)=G(x,0,t)$ \cite{baskin}. 
Therefore, the mean squared displacement
of mobile particles is anomalous subdiffusion:
\be\label{c10} 
\lgl x^2(t)\rgl=\frac{\lgl x^2(t)c(x,t)\rgl}{\lgl c(x,t)\rgl}= D_1(\pi
t/D_2)^{1/2} \, . 
\ee 
This mean squared displacement $\lgl x^2(t)\rgl$ is a mean radial size of the tumor 
after elapsing transport time $t$. 
For the TTT model it means that a general elapsed time is
\be\label{c11}
t\equiv t_N=\sum_{i=1}^N(\cT_f(i)+\cT_t(i))=t_f+t_t.
\ee
During the fission time $t_f$ neoplasm cells do not move, then  
 $t=t_t$ in (\ref{c10}). 
If $t_t$ is large enough, this process may give a possible explanation not
only for the mean size of the primary tumor but also  it could be  
mean distance between the primary tumor and a metastasis.

 The transport (or grow) process accelerated by including so-called 
L\'evy--type flights into consideration. In the framework of the comb 
model it still leads to subdiffusion with transport exponent $ 
1/2<\alpha\leq 1$ \cite{basiom}. Therefore to obtain a superdiffusion 
process one needs to go beyond the comb model consideration.
The transport can be superdiffusive 
due to Levy-type flights when cells may traverse
all of the lymph nodes ultimately to reach the vascular compartment via the
thoratic duct. 
In this case the Levy--type process  is a result of the cell transport 
in the vascular system.
Therefore, the solution of Eq.
(\ref{c9}) should be modified in the following way: along the
structure axis we add L\'{e}vy flights with a L\'{e}vy
distribution for the jump length with the asymptotic behavior
\cite{klafter} 
\be\label{c12}
 \tilde{\psi}(x)\sim |x|^{-1-\mu}\, ,
\ee
where $1<\mu<2$.
The solution Eq. (\ref{c9}) is modified, for example,
in the following form \cite{add1}
\be\label{c13}
G_L(k,0,p)=\frac{1}{D_2p^{1/2}+D_1|k|^{\mu}} \, .
\ee
The power--law asymptotics for the Green function is
\be\label{c14}
c(x,t)\sim \frac{t^{-g_{\mu}}}{|x|^{1+\mu}} \, 
\ee
with $g_{\mu} > 0$. For example, when $\mu = 1$, $g_{\mu} =1$.
In this case the mean squared displacement diverges:
\be\label{c15}
\lgl x^2(t)\rgl\longrightarrow\infty \, .
\ee
This expression of Eq. (\ref{c15})  could be a simple explanation
of appearance of metaphases at any long distance from the primary tumor.

\section{Conclusion}

A cancer growth process is considered in the framework of the TTT model. 
It is shown that the main process of the tumor development is a fractional 
transport of cancer cells on the lymphatic net. In the framework of this 
fractional
kinetics we were able to show that the mean size of the main growth is due 
to subdiffusion, while the appearance of metaphases is determined by
superdiffusion. The underlying mechanism of the tumor development is a 
partition
of this grow on fission times $\cT_f(i)$ and transport times $\cT_t(i)$.
Due to this assumption, there are free cancer cells, which contributes
to fractional transport. Therefore, both the characteristic size
of the cancer growth and metaphases are related to the fractional
transport exponents $\alpha,\mu$, which depends on geometrical
fractional dimension of the lymphatic net $d_f$.

{\bf Acknowledgment}

We thank E. Baskin for helpful discussions and critical remarks.


\begin{thebibliography}{99}
\bibitem{add1} This question is not simple and deserved a separate 
consideration \cite{basiom}.
\end{thebibliography}
\end{document}